\documentstyle[preprint,aps,epsfig]{revtex}
\draft
\begin{document}
\preprint{
\begin{tabular}{r}
SNUTP 96-059
\\
KAIST-TP 96/10
\\
hep-ph/yymmdd
\end{tabular}
}
\title{Medium Effects on the CP phases  and Dynamical Mixing
Angles in the Neutrino Mixing Matrix}
\author{
J. D. Kim$^{\mathrm{a}}$\thanks{E-mail address: jdkim@phyb.snu.ac.kr},
C. W. Kim$^{\mathrm{b,c}}$\thanks{E-mail address: cwkim@jhuvms.hcf.jhu.edu},
Jewan Kim$^{\mathrm{a,d}}$\thanks{E-mail address: jewan@jewan.snu.ac.kr},
and
Dae-Gyu Lee$^{\mathrm{a}}$\thanks{E-mail address: leedaegy@ctp.snu.ac.kr}
}
\address{
\begin{tabular}{c}
$^{\mathrm{a}}$Department of Physics, Seoul National University, Seoul, Korea
\\
$^{\mathrm{b}}$Department of Physics, Korea Advanced Institute of Science
and Technology, Daeduk, Korea 
\\
$^{\mathrm{c}}$Department of Physics and Astronomy, The Johns Hopkins University, 
Baltimore, Maryland 21218\thanks{Permanent Address}
\\
$^{\mathrm{d}}$Center for Theoretical Physics, Department of Physics,
Seoul National University, Seoul, Korea
\end{tabular}
}
\date{\today}
\maketitle
\begin{abstract}
The concepts of effective and dynamical neutrino mixing matrices
are introduced in order to describe the behavior of neutrinos in matter. 
The former relates weak eigenstates 
to mass eigenstates, whereas the latter relates weak eigenstates
to energy eigenstates in matter. It is shown that the  
dynamical mixing angles enable us to express the neutrino
survival probability in the Sun without any resort to the 
Landau-Zener transition probability for the non-adiabatic process.
Also discussed are effective CP violating phases that appear
in the effective and dynamical mixing matrices in matter.
Both two and three generation cases are discussed
using the solar neutrinos as an example.
\end{abstract}

\narrowtext

\section{Introduction}
The Mikheyev-Smirnov-Wolfenstein 
(MSW)~\cite{MSW1,MSW2} effect has been 
considered for some time  as the most plausible mechanism to explain
the solar neutrino puzzles~\cite{HOMESTAKE,KAMIOKANDE,GALLEX,SAGE}. 
This effect dramatically enhances the conversion of a flavor
neutrino ($\nu_{e}$ in the case of solar neutrinos) into another
when certain conditions among the intrinsic neutrino properties and
the nature of matter are met in a medium.
When the matter density in the Sun changes slowly enough
(to be determined by the intrinsic properties 
such as masses and mixing angles and by the neutrino energy) so that
the neutrino conversion from one flavor to another takes place
adiabatically, the MSW effects are
known to be well described by 
the effective masses and mixing angles in matter. 
These effective
values at any given time or distance from the production in matter
are commonly derived from the equation of motion for the weak
eigenstates in matter by a simple matrix diagonalization
under the assumption that the matter density is uniform.

However, when the time  (or distance) variation of the density
 in matter is sufficiently rapid, transitions among the effective mass
eigenstates take place and the process becomes non-adiabatic.
In this case, the role of the effective values becomes vague.
In order to overcome this, the neutrino survival probability is 
approximately expressed in terms of the Landau-Zener transition
probability~\cite{CROSS1,CROSS2,LANDAU,ZENER,THREE1,THREE2,THREE3,THREE4}. 
The other alternative is, of course, to solve numerically
the equation of motion for the weak eigenstates in matter. 
It has been recognized, however, that to do so is prohibitively
difficult when all three generations of the neutrinos are involved.

In this paper,we first introduce two concepts of the
mixing angles in matter, effective and dynamical
mixing angles and the corresponding mixing matrices. 
It will be shown that the dynamical mixing angles enable
us to calculate the survival probability without resort to
the Landau-Zener transition probability. The effective mixing
angles defined here are equivalent to those that have been
commonly  used in the past, i.e., they appear in the effective mixing
matrix which relates the weak eigenstates 
and the mass eigenstates in matter.
As mentioned already, these angles are sufficient to describe the adiabatic
process and have also been used  for the non-adiabatic  case together with
the Landau-Zener transition probability.

The dynamical mixing angles are defined to appear in the mixing matrix
which relates the weak eigenstates and the {\it energy} eigenstates in matter.
(Note that even in the case when there appear transitions 
among the mass eigenstates,
the energy eigenstates are well defined.)
It will be shown that the neutrino survival probability can be
 expressed in terms of
the dynamical mixing angle alone even 
in the case of non-adiabatic processes. 
Since in this approach no approximations are involved in calculating
the survival probability, the results are exact whereas 
the usual treatment with the use of the Landau-Zener transition probability
relies on some approximations.

The second subject to be discussed in this paper is the 
medium effect on the CP violating phases 
already present in the neutrino  mixing matrix.
In the case of general mixings,  it has been known that the CP phases
would not significantly modify the neutrino transition probabilities,
in particular when the oscillating terms are averaged out. Therefore,
an observation of the CP violating effects is rather difficult, if
not impossible. This general conclusion, however, needs not to be 
correct in the case of maximal mixings.
The possible solution of the atmospheric neutrinos and the so-called
large angle solution of the solar neutrino problem require  rather
large mixing angles, suggesting the possibility of maximal mixings
of three neutrinos~\cite{MAX3,MAX5}. 
In this case, it is necessary to have a CP
violating phase. That is, maximal mixings are impossible without
CP violating phases~\cite{MAX4}.

In this paper, we investigate how the medium effects would modify
the CP violating phases in the mixing matrix in the vacuum. 
It will be shown that in the case of three generations of neutrinos,
in addition to numerical  changes 
in the original vacuum CP phase, the medium effects
induce additional {\it effective} CP phases which, of course,
would  vanish when 
the matter effects are turned off. Although it is necessary to have
additional CP phases in matter, they  are not independent of the 
original vacuum CP phases. 
In the two generation case, no additional CP phase appears in matter.

This paper has been organized as follows: In Section II 
we  introduce  two mixing matrices, the effective mixing matrix
and the dynamical mixing matrix. 
Section III is devoted to a detail discussion of
the effective mixing matrices in two and three generations of
neutrinos. In particular, we demonstrate how additional induced
phases appear in a medium in the case of three generations.
In Section IV, we discuss the dynamical mixing matrices.
After presenting a simple example of the two generation case,
we discuss the case of three generations with non-zero CP.
We summarize our results and conclude in Section V.

\section{Definitions of Effective  and Dynamical mixing matrices}
Let us begin with the  wave equation
for neutrinos propagating through a medium
with a time-dependent potential, $V(t)$,
being felt by the neutrinos {\it a la} MSW effects
\begin{equation}
i\frac{d}{dt}|\Psi_w(t)> = \frac{1}{2E}M^w|\Psi_w(t)> ,
\label{sch}
\end{equation}
where $|\Psi_w(t)>$ is the neutrino wave function in 
the weak (flavor) basis
and $M^w$ is the time evolution matrix in the weak basis
which dictates the motion of the weak eigenstate neutrinos.
First, we define the {\it effective mixing matrix} $U_{eff}$ which 
satisfies the following equation
\begin{equation}
  U^{\dagger}_{eff}M^wU_{eff} = M^m , 
\label{eff}
\end{equation}
where $M^m$ is the diagonal mass matrix in the mass basis. 
Solving the neutrino wave equation Eq.~(\ref{sch}) is
equivalent to finding the effective mixing matrix $U_{eff}$  in Eq.~(\ref{eff}) 
in the case where neutrinos propagate
in a medium with a uniform density ($V(t)=$ constant).
 The result is sufficient to
describe the adiabatic MSW process in a medium.

Next, we introduce a unitary matrix $U$ which diagonalizes 
the time-evolution matrix for a time-dependent $V(t)$.
By transforming the wave function in the weak basis 
with the unitary matrix $U$, we obtain 
the wave function in the energy basis such that its time
evolution is of  the form
\begin{equation}
 |\Psi_e(t)> = e^{-i\frac{1}{2E}\int^t_0M^e(t)dt} |\Psi_e(0)>  ,
\end{equation} 
where $M^e(t)$ is diagonal and 
the subscript $e$ in the wave function $\Psi_e$ 
denotes the energy eigenstate. 
In other words, the above-introduced unitary matrix $U$ 
relates 
$ |\Psi_w(t)>$ and $ |\Psi_e(0)>$ 
for a time-dependent $V(t)$ as 
\begin{equation}
 |\Psi_w(t)> = U(t)|\Psi_e(t)>    .
\label{tran}
\end{equation}
By inserting Eq.~(\ref{tran}) into Eq.~(\ref{sch}), we can obtain 
a differential equation for the matrix $U$. 
From now on , we will call this $U$
the {\it dynamical mixing matrix}
and use the notation $U_{dyn}$.
The  $U_{dyn}$ satisfies the differential equation
\begin{equation}
  U^{\dagger}_{dyn}M^wU_{dyn} = M^e + 2EiU^{\dagger}_{dyn}\dot{U}_{dyn} ,
\label{dyn}
\end{equation}
where $M^e$ is a diagonal matrix 
whose elements are composed of time varying
energies of the Hamiltonian $M^w/2E$.

The form of the dynamical mixing matrix $U_{dyn}$ 
does not depend on whether the neutrinos are
Dirac or Majorana type.
For the two generation case, the dynamical 
mixing matrix is a $2\times 2$
matrix with one dynamical mixing angle and 
one dynamical phase if the effective potential $V(t)$ is not a constant
in time. 
For the three generation case, the dynamical mixing matrix is 
a $3\times 3$ matrix with
three dynamical mixing angles and three dynamical phases.

Summarizing the above,
the effective mixing matrix relates the 
weak eigenstates and the mass eigenstates
\begin{equation}
  |\hbox{ weak eigenstates}> = U_{eff}|\hbox{mass eigenstates}>,
\end{equation}
whereas
the dynamical mixing matrix relates 
 the weak eigenstates and the energy eigenstates
\begin{equation}
  |\hbox{ weak eigenstates}> = U_{dyn}|\hbox{energy eigenstates}>.
\end{equation}
Usually, it is easy to calculate
the effective mixing matrix  by using simple algebra 
, i.e.,
it is an eigenvalue and eigenvector problem. 
The eigenvalues correspond to
the effective masses and an appropriate assemble of eigenvectors 
is nothing but the
effective mixing matrix defined by Eq.~(\ref{eff}).
As for the dynamical mixing matrix, however, 
it is non-trivial to obtain the
dynamical mixing angles, phases and 
the time-dependent energy eigenvalues 
because it is equivalent to solving  the Schr\"{o}dinger 
equation of the neutrino wave function given in Eq.~(\ref{sch}).
Therefore, one often resorts to numerical methods to
solve Eq.(1). 

\section{Effective neutrino mixing matrices}
\subsection{Two Generation Case}

 As a simple example, 
 let us consider the two generation mixing
of Majorana neutrinos to investigate how the CP phase in
vacuum would be modified in matter.
(In the case of Dirac neutrinos, the phase is trivial, that is, zero
in matter as well   as in vacuum.) 
In general, the mixing matrix in vacuum
can be written as
\begin{equation}
  U_0 = \left(
    \begin{array}{cc}
      \cos\theta   &  \sin\theta e^{i\omega}  \\
     -\sin\theta e^{-i\omega}  & \cos\theta
    \end{array}  \right).
\end{equation}
In Eq.(8) and hereafter,
a quantity with subscript $0$ always denotes 
its value in vacuum.
The mass matrix in  the weak basis can be obtained from the mass matrix 
in the mass basis as
\begin{equation}
  M^w_0 = U_0M^m_0U^{\dagger}_0 \equiv \left(
    \begin{array}{cc}
      m_{11}  &  m_{12}   \\
      m_{21}  &  m_{22}
    \end{array}   \right)          ,
\end{equation}
where $M^m_0=\hbox{diag}(m^2_1,m^2_2)$.

In matter, a quantity $A(t)\equiv 2E_{\nu}V(t)$, which is 
the amount of the increase 
in the effective mass squared in matter for $\nu_e$, 
is inserted in the 1-1 element of the mass matrix 
\begin{equation}
  M^w = \left(
    \begin{array}{cc}
      m_{11}+A  &  m_{12}   \\
      m_{21}  &  m_{22}
    \end{array}   \right)
\label{mw1}
\end{equation}
with $A=2E_{\nu}V=2\sqrt{2}G_FE_{\nu}N_e$.
(We ignore the contributions from neutral current interactions.)
The $G_F$, $E_{\nu}$, and $N_e$
represent, respectively, the Fermi coupling constant,
the energy of the neutrino under consideration,
and the electron number density in a medium.
Let us denote the effective mixing matrix 
in matter as 
\begin{equation}
  U_{eff} = \left(
    \begin{array}{cc}
      \cos\bar{\theta}   &  \sin\bar{\theta} e^{i\bar{\omega}}  \\
     -\sin\bar{\theta} e^{-i\bar{\omega}}  & \cos\bar{\theta}
    \end{array}  \right).
\end{equation}
This effective mixing matrix $U_{eff}$ diagonalizes the effective mass matrix
in the weak basis,i.e., 
$U_{eff}^{\dagger}M^wU_{eff} = M^m$,
where $M^m=\hbox{diag}(\mu^2_1,\mu^2_2)$.
The effective masses $\mu_1,\mu_2$ and the effective mixing
angle $\bar{\theta}$ are well known and are
given by (For recent reviews, see for example, \cite{THREE1,CWKIM}) 
\begin{eqnarray}
\mu^2_1  &=& \frac{m^2_1+m^2_2}{2}+\frac{A}{2}
   -\frac{1}{2}\sqrt{(A-\Delta_0^2\cos2\theta)^2+\Delta_0^4\sin^22\theta} ,\\
\mu^2_2  &=& \frac{m^2_1+m^2_2}{2}+\frac{A}{2}
   +\frac{1}{2}\sqrt{(A-\Delta_0^2\cos2\theta)^2+\Delta_0^4\sin^22\theta} , \\
\tan2\bar{\theta} &=& \frac{\tan2\theta}{1-A/\Delta_0^2\cos2\theta}  ,
\end{eqnarray}
where $\Delta_0^2\equiv m^2_2-m^2_1$.
In the two generation case, 
a simple algebra shows that the effective phase $\bar{\omega}$ remains
unaffected by matter, i.e., $\bar{\omega}=\omega$.
However, this is not the case for the three generation case as well as 
for the two generation case of the dynamical mixixing matrix, 
as we will see later.
The above information is sufficient to describe the adiabatic
conversion processes of $\nu_e$ in the Sun.

\subsection{Three Generation Case}
Now let us consider the case of three generations of neutrinos
by generalizing the case discussed above.
The representation of the unitary mixing matrix~\cite{MAIANI,CKM1,CKM2}
depends on the neutrino type. 
Majorana neutrinos can have three non-zero
CP phases in vacuum whereas Dirac neutrinos can have only
one CP phase.
For simplicity, we will consider the case where only one
non-zero CP phase appears in vacuum. 
The result which we will present below can easily be
generalized to the case with more than one vacuum CP phases. 
 
We have found it convenient to describe the neutrino mixing
with the use of the modified Maiani representation 
as advocated in Particle Data Group~\cite{PDB} 
given by
\begin{equation}
U_0=\left(
     \begin{array}{ccc}
   c_{12}c_{13}         & s_{12}c_{13}   & s_{13}e^{-i\delta} \\
  -s_{12}c_{23}- c_{12}s_{13}s_{23}e^{i\delta} 
 & c_{12}c_{23}- s_{12}s_{13}s_{23}e^{i\delta} & c_{13}s_{23}  \\
   s_{12}s_{23}- c_{12}s_{13}c_{23}e^{i\delta} 
 & -c_{12}s_{23}- s_{12}s_{13}c_{23}e^{i\delta} & c_{13}c_{23}  
     \end{array}
  \right)     ,
\label{vacu}
\end{equation}
where $s_{ij}\equiv\sin\theta_{ij}$,
$c_{ij}\equiv\cos\theta_{ij}$ .
(The mixing angles $\theta_{ij}$ and 
phase $\delta$ in Eq.~(\ref{vacu}) are all vacuum values
although they do not have the subscript $0$.)

The mass matrix in the weak basis is related to 
the mass matrix in the mass basis as 
\begin{equation}
  M^w_0 = U_0 M^m_0 U^{\dagger}_0 \equiv\left(
 \begin{array}{ccc}
         m_{11}  & m_{12} & m_{13}  \\
         m_{21}  & m_{22} & m_{23}  \\
         m_{31}  & m_{32} & m_{33}
 \end{array} \right)      ,
\end{equation}
where $M^m_0=\hbox{diag}(m^2_1,m^2_2,m^2_3)$.
In a way analogous to the two generation case,
the matter effect is inserted in the mass matrix: 
\begin{equation}
   M^w=\left(
       \begin{array}{ccc}
         m_{11} + A & m_{12} & m_{13}  \\
         m_{21}     & m_{22} & m_{23}  \\
         m_{31}     & m_{32} & m_{33}
       \end{array} \right)  . 
\end{equation}
Now, solving the problem in matter amounts to diagonalizing 
the Hamiltonian or solving the time evolution matrix in Eq.~(\ref{sch}).
In the adiabatic case or in the case of a uniform matter density,
it is equivalent to finding the effective mass matrix $M^m$
and the effective mixing matrix $U_{eff}$ from the equation
$U_{eff}^{\dagger}M^wU_{eff}=M^m$, 
where  $M^m=\hbox{diag}(\mu^2_1,\mu^2_2,\mu^2_3)$.
The eigenvalues are
\begin{eqnarray}
\mu^2_1 & = & 2\alpha^{1/3}\cos{(\frac{\beta+2\pi}{3})}-\frac{b}{3} ,\\
\mu^2_2 & = & 2\alpha^{1/3}\cos{(\frac{\beta-2\pi}{3})}-\frac{b}{3} ,\\
\mu^2_3 & = & 2\alpha^{1/3}\cos{(\frac{\beta}{3})}-\frac{b}{3}      ,
\end{eqnarray}
where
\[
   \alpha\equiv\sqrt{q^2 + |p^3 + q^2|} , \hspace{1cm}
   \beta \equiv\arctan{(\frac{\sqrt{|p^3+q^2|}}{q})}     ,
\]
and
\[
p  =  \frac{c}{3}-\frac{b^2}{9}, \hspace{2cm} 
q  =  -\frac{d}{2} + \frac{bc}{6}-\frac{b^3}{27}      ,
\]
with
$b=-\hbox{Tr}M^w$, $c=\frac{1}{2}[(\hbox{Tr}M^w)^2-\hbox{Tr}(M^w)^2]$,
and $d=-\hbox{det}M^w$.

Let us denote the eigenvector corresponding to the third eigenvalue
as a column vector $(c_1,c_2,c_3)^T$, where $T$ denotes 
transpose.
Solving the equations for $c_1,c_2$ and $c_3$,
we obtain the ratios
\begin{equation}
\frac{c_2}{c_1}=\frac{(m_{33}-\mu_3^2)m_{21}
     -m_{23}m_{31}}{m_{23}m_{32}-(m_{22}-\mu_3^2)(m_{33}-\mu_3^2)}  ,
\end{equation}
\begin{equation}
\frac{c_3}{c_1}=\frac{(m_{22}-\mu_3^2)m_{31}
     -m_{32}m_{21}}{m_{23}m_{32}-(m_{22}-\mu_3^2)(m_{33}-\mu_3^2)}.
\end{equation}
In matter, the phases of $c_2/c_1$ and $c_3/c_1$ are no longer the same 
due to the presence of the non-zero vacuum CP phase $\delta$.
This is the reason why we are forced to introduce
two additional phases for the mixing matrix 
to be unitary and self-consistent.
Thus, we can not use the same form as the modified Maiani representation 
given in Eq.~(\ref{vacu}) in order to describe 
the new effective matrix in matter.

Let us take a representation in matter which has 
two additional {\it induced} phases, $\xi$ 
and $\eta$ which appear with $\theta_{23}$ and $\theta_{12}$, respectively,
\begin{equation}
U_{eff}=\left(
     \begin{array}{ccc}
   \bar{c}_{12}\bar{c}_{13}             & 
   \bar{s}_{12}\bar{c}_{13}e^{i\eta}    & 
   \bar{s}_{13}e^{-i\bar{\delta}}      \\
  -\bar{s}_{12}\bar{c}_{23}e^{-i\eta}
  -\bar{c}_{12}\bar{s}_{13}\bar{s}_{23}e^{i(\bar{\delta}-\xi)} &
   \bar{c}_{12}\bar{c}_{23}
  -\bar{s}_{12}\bar{s}_{13}\bar{s}_{23}e^{i(\bar{\delta}+\eta-\xi)} & 
   \bar{c}_{13}\bar{s}_{23}e^{-i\xi}  \\
   \bar{s}_{12}\bar{s}_{23}e^{-i(\eta-\xi)}
  -\bar{c}_{12}\bar{s}_{13}\bar{c}_{23}e^{i\bar{\delta}} & 
  -\bar{c}_{12}\bar{s}_{23}e^{i\xi}
  -\bar{s}_{12}\bar{s}_{13}\bar{c}_{23}e^{i(\bar{\delta}+\eta)} & 
   \bar{c}_{13}\bar{c}_{23}
     \end{array}
  \right).
\label{new}
\end{equation}
where $\bar{s}_{ij}\equiv\sin{\bar{\theta}_{ij}}$,
$\bar{c}_{ij}\equiv\cos{\bar{\theta}_{ij}}$. 
A barred quantity  denotes its effective value in matter.
The two additional phases, $\xi$ and $\eta$, are 
{\it induced} in matter due to the presence of the original non-zero CP phase
in vacuum.
With the representation Eq.~(\ref{new}),
we can obtain all the effective values in matter.

The two effective mixing phases are given by
\begin{eqnarray}
\tan\bar{\delta} &=& \frac{\hbox{Im}(c_3/c_1)}{\mbox{Re}(c_3/c_1)}  
\label{delta} , \\
\tan(\bar{\delta}-\xi) &=& \frac{\hbox{Im}(c_2/c_1)}{\mbox{Re}(c_2/c_1)},
\end{eqnarray}
and the effective mixing angles are
\begin{eqnarray}
\tan\bar{\theta}_{13} &=& \frac{1}{\sqrt{|c_2/c_1|^2+|c_3/c_1|^2}} ,   \\
\cos\bar{\theta}_{23} &=& \frac{|c_3/c_1|}{\sqrt{|c_2/c_1|^2+|c_3/c_1|^2}} .
\end{eqnarray}
In order to obtain the remaining mixing angle $\bar{\theta}_{12}$ and the phase $\eta$,
we must solve another eigenvector
for an eigenvalue, for example, $\mu_1^2$. 
Let us denote the eigenvector as
a column vector $(d_1,d_2,d_3)^T$.
Solving the equations for $d_1,d_2$ and $d_3$,
we obtain the ratio of $d_2$ to $d_1$
\begin{equation}
\frac{d_2}{d_1}=\frac{(m_{33}-\mu_1^2)m_{21}-
         m_{23}m_{31}}{m_{23}m_{32}-(m_{22}-\mu_1^2)(m_{33}-\mu_1^2)}.
\end{equation}
After comparing with the ratios of 
the 1-1 and 1-2 mixing matrix elements,
we obtain the effective phase $\eta$ and 
the effective mixing angle $\bar{\theta}_{12}$ as
\begin{eqnarray}
\tan\eta  &=& -\frac{\tan\bar{\theta}_{13}\sin\bar{\theta}_{23}
         \sin(\bar{\delta}-\xi)
      +\mbox{Im}(d_2/d_1)}{\tan\bar{\theta}_{13}\sin\bar{\theta}_{23}
         \cos(\bar{\delta} -\xi)+\mbox{Re}(d_2/d_1)}  , \label{eta} \\
\tan\bar{\theta}_{12} &=& -(\tan\bar{\theta}_{13}
         \sin\bar{\theta}_{23}\cos(\bar{\delta}-\xi)+ \mbox{Re}(d_2/d_1))
     \frac{\cos\bar{\theta}_{13}}{\cos\bar{\theta}_{23}\cos\eta}.
\end{eqnarray}
From Eqs.~(\ref{delta}) and (\ref{eta}), we can see that the two
additional induced phases $\xi$ and $\eta$ are generated by the non-zero vacuum CP 
phase $\delta$. That is, if we set $\delta=0$, then
Eq.~(\ref{delta}) shows that the effective phase $\bar{\delta}$
also vanishes 
because $c_3/c_1$ cannot have an imaginary part. Similarly, 
it is easy to see that
the two additional phases $\xi$ and $\eta$ vanish if $\delta=0$.
We emphasize that the  
two phases $\xi$ and $\eta$ are not new and independent phases,
but instead they are related to the non-zero vacuum CP phase $\delta$.

In order to exhibit this property quantitatively, we use 
the solar neutrinos as an example.
Most solar neutrinos are produced in the region
$x\leq 0.25$ 
where $x$ is the ratio of
the distance $R$ from the center of the Sun and the radius of the Sun
$R_{\odot}$. 
In the Sun,
the electron number density distribution~\cite{BAH2,BAH1} 
is well fitted by              
$\log_{10}(N_e(x)/N_A)=2.32-4.17x-0.000125/[x^2+0.5^2]$
where $N_A$ is Avogadro's number per cm$^3$. 
The quantity $A$ in the center of the Sun is
$A\cong 3.17\times 10^{-5}(E_{\nu}/\hbox{MeV})$eV$^2$.
With the solar neutrino energy $E_{\nu}\leq 14$MeV,
the largest value of $A$ in the Sun is of the order of 
$10^{-3}$eV$^2$. Therefore, if we assume small vacuum mixing angles,
there are two resonance points ($\bar{\theta}_{ij}=\pi/4$) 
inside the Sun
when the mass squared difference $\Delta_{31}=m^2_3-m^2_1$ is less 
than $10^{-3}$eV$^2$. 

Fig. 1 and Fig. 2 show the typical behavior of
the three effective mixing angles and three phases
in the case where the neutrino mass squared
differences are taken as $\Delta_{21}\sim 10^{-4}$eV$^2$,
$\Delta_{31}\sim 10^{-2}$eV$^2$.
In Fig. 1, we have taken
the vacuum mixing angles 
$\theta_{12}=\theta_{13}=\theta_{23}=0.05$
and the vacuum CP phase $\delta=0.5$.  
Therefore, we can see that, in contrast to the case of the effective
mixing angles and effective masses which are dramatically affected
by the medium, the CP phase $\bar{\delta}$ remains unchanged, 
and $\xi$ and $\eta$ are induced for large values of $A$
in the adiabatic processes.
Fig. 2 is for the vacuum mixing angles
$\theta_{12}=\theta_{23}=\pi/4$, $\theta_{13}=0.6155$,
and the vacuum CP phase $\delta=\pi/2$.
This is the maximal mixing case. In Fig. 2,
the effective mixing angles $\bar{\theta}_{12}$ and 
$\bar{\theta}_{13}$ change from
$\theta_{12}=\pi/4$ and  $\theta_{13}=0.6155$  
in vacuum to 
$\bar{\theta}_{12}\sim\bar{\theta}_{13}\sim\pi/2$ 
in the solar interior, whereas the mixing angle
$\bar{\theta}_{23}$ and the phase $\bar{\delta}$ also
practically remain unchanged. 
Again, the induced $\xi$ and $\eta$ appear when $A$ becomes large.
This confirms the assertion made in \cite{MAX1,MAX2}
that the matter effects are practically 
absent in the maximal mixing case.

In general, the non-zero CP phase contributes to the transition probability
among the flavor neutrino states. For simplicity, we consider
the averaged survival probability of $\nu_e$ in the adiabatic approximation.
The survival probability is given by
\begin{eqnarray}
<P(\nu_e\rightarrow\nu_e;\hbox{adiabatic})>  &=&  \nonumber \\  
  & & \hspace{-3cm} (\bar{c}_{12}\bar{c}_{13}c_{12}c_{13})^2
     +(\bar{s}_{12}\bar{c}_{13}s_{12}c_{13})^2
                           +(\bar{s}_{13}s_{13})^2    \nonumber  \\
  & &\hspace{-3cm} +2\bar{c}_{12}\bar{c}_{13}\bar{s}_{12}\bar{c}_{13}
           c_{12}c_{13}s_{12}c_{13}\cos\eta            \nonumber  \\
  & &\hspace{-3cm} +2\bar{c}_{12}\bar{c}_{13}\bar{s}_{13}
           c_{12}c_{13}s_{13}\cos(\bar{\delta}-\delta)    \nonumber \\
  & &\hspace{-3cm} +2\bar{s}_{12}\bar{c}_{13}\bar{s}_{13}
           s_{12}c_{13}s_{13}\cos(\eta+\bar{\delta}-\delta).
\label{prob}
\end{eqnarray}
Although three phases are present in Eq.~(\ref{prob}), 
since they appear as
$\cos\eta,\cos(\bar{\delta}-\eta)$, and
$\cos(\eta+\bar{\delta}-\delta)$, 
all of which are close to unity,
the CP phases are almost impossible to detect in practice.
(Note in the figure that $\bar{\delta}\sim\delta$
and $\xi,\eta\sim 0$).

\section{Dynamical neutrino mixing matrices}
\subsection{Two Generation Case}
Now let us study the dynamical mixing matrix in detail. 
We will first consider the two generation case. 
From the definition of the dynamical
mixing matrix given by Eq.~(\ref{dyn}), we can write 
the differential matrix equation in a
$2\times 2$ matrix form.
As it becomes clear later, at least one non-zero phase is necessary independently 
of the neutrino type even in the two generation case. 
With the following notations
\begin{eqnarray}
 U_{dyn} &\equiv& \left(
         \begin{array}{cc}
          \cos\psi               & \sin\psi e^{i\alpha} \\
          -\sin\psi e^{-i\alpha} & \cos\psi
         \end{array} \right), \\
 M^e &\equiv & \left(
         \begin{array}{cc}
           {\cal E}_1   &   0   \\
            0                & {\cal E}_2
         \end{array}  \right) ,
\end{eqnarray}
and
\begin{equation} 
 M^w = \left(
         \begin{array}{cc}
 -\frac{\Delta_0^2}{2}\cos{2\theta}+A(t)  & \frac{\Delta_0^2}{2}\sin{2\theta}    \\
 \frac{\Delta_0^2}{2}\sin{2\theta}     & \frac{\Delta_0^2}{2}\cos{2\theta} 
         \end{array}  \right) ,
\label{mw2}
\end{equation}
Eq.~(\ref{dyn}) leads to the two coupled equations for $\psi$ and $\alpha$
\begin{eqnarray}
 \dot{\psi}   &=&  -\frac{\Delta_0^2}{4E}\sin{2\theta}\sin\alpha  ,
\label{psi2}                    \\
 \dot{\alpha} &=& -\frac{\Delta_0^2}{2E}\sin{2\theta}\cos\alpha\cot 2\psi 
                     +\frac{\Delta_0^2\cos{2\theta}-A(t)}{2E}     .
\label{alpha2}                                 
\end{eqnarray}
Here, it is easy to see that if there is no phase (i.e., $\alpha=0$), 
it is impossible to describe the case where $A$ changes in time
since with $\alpha=0$ Eq.~(\ref{alpha2}) is valid only at a fixed
value of $t$.
In the above, the mass matrix in the weak basis in Eq.~(\ref{mw2}) has been obtained 
from the mass matrix Eq.~(\ref{mw1}) by subtracting 
$(m^2_1+m^2_2)/2$ from the diagonal elements in Eq.~(\ref{mw1}).
In addition, we obtain
the energy eigenvalues, ${\cal E}_1(t)$ and ${\cal E}_2(t)$,
\begin{eqnarray}
 {\cal E}_1(t) &=& -\frac{\Delta_0^2}{2}\cos{2\theta}+A(t)
        -\frac{\Delta_0^2}{2}\sin{2\theta}\cos\alpha\tan\psi , \\
 {\cal E}_2(t) &=& \frac{\Delta_0^2}{2}\cos{2\theta}
       -\frac{\Delta_0^2}{2}\sin{2\theta}\cos\alpha\tan\psi .
\end{eqnarray}
Equations~(\ref{psi2}) and (\ref{alpha2}) are coupled 
non-linear differential equations. 
Similar equations have previously been obtained in \cite{LK}
as a means to describe the non-adiabatic process without
the use of the Landau-Zener transition probability.
The initial conditions for $\psi$ and $\alpha$ are fixed by the vacuum values.
Given $A(t)$, we can solve the equations numerically. 
For $E/(m^2_2-m^2_1)\rightarrow\infty$, $\psi(t)$ and $\alpha(t)$ approach
the vacuum values, which can be seen from the usual treatment 
with the Landau-Zener formula.
Fig. 3 (a) (Fig. 3 (b)) shows
the dynamical mixing angle and phase in the case of
the neutrino energy $E_{\nu}=10$ MeV, 
the vacuum mixing angle $\theta=0.1$, and
$\Delta^2_0=10^{-4}\hbox{eV}^2$ 
($\Delta^2_0=10^{-6}\hbox{eV}^2$).
The difference between the effective and dynamical mixing angles
depends sensitively on $\Delta^2_0$. 
Fig. 3 (c) shows the dynamical mixing angle
and phase in the maximal mixing case of two neutrinos 
with the neutrino energy $E_{\nu}=10$ MeV
and the mass squared difference $\Delta^2_0=10^{-6}\hbox{eV}^2$.
The abscissa $x$ indicates the ratio of the distance 
of interest from the surface of the Sun
to the solar radius, i.e., $x=0$ corresponds to the surface
of the Sun and $x=1$ the center of the Sun.
The sign of the dynamical phase $\alpha$ depends on the direction of 
the neutrino motion, i.e., the sign of $\alpha$ is reversed
as $\nu_e$ travels inwards from the solar surface.
The behavior of $\alpha$ as a function of $x$ is quite remarkable 
as $x\rightarrow 1$.

The time averaged survival probability for $\nu_e$
is given by
\begin{equation}
<P(\nu_e\rightarrow\nu_e)>=\cos^2\psi\cos^2\theta+
                           \sin^2\psi\sin^2\theta       .
\label{tran2}
\end{equation}
We emphasize here that although the expression in Eq.~(\ref{tran2})
is identical to the usual survival probability 
for the adiabatic process when $\psi$ is replaced by
the effective mixing angle $\bar{\theta}$,
it is an exact one which can describe 
the non-adiabatic process without using the Landau-Zener transition 
probability.
In Eq.~(\ref{tran2}) the dynamical phase $\alpha$
has disappeared in the process of time average.
Fig. 4  shows the survival probabilities
given by Eq.~(\ref{tran2})
as functions of $E_{\nu}/\Delta^2_0$ for three examples with
the vacuum mixing angles $\theta=0.01$,  $\theta=0.1$
and $\theta=\pi/4$ (maximal mixing in two generations).
The transition probability for the maximal mixing case is
practically constant at 0.5, in the region of
$10^{-10}\leq E_{\nu}/\Delta_0^2\leq 10^{-15}$ and 
the neutrino oscillation in matter cannot 
be differentiated from the pure vacuum oscillation.
However it does not mean that 
the dynamical mixing angle $\psi$ remains constant in matter.
The value of the angle $\psi$ approaches to $\pi/2$ 
near the center of the Sun.
In Fig. 4 the survival probabilities
calculated using the Landau-Zener transition probability (dotted lines
for $\theta=0.01, \theta=0.1$)
are compared with the exact results
based on the use of dynamical mixing angle.
\subsection{Three Generation Case}
In the case of three generations, there are six coupled
differential equations and three relations to be solved.
The unknown variables are three dynamical
mixing angles, three dynamical phases and three energy eigenvalues.
Let us write the dynamical neutrino mixing matrix 
in the three generation as
\begin{equation}
U_{dyn}=\left(
\begin{array}{ccc}
1      &      0              &        0               \\
0      &   c_{23}            &   s_{23}e^{-i\gamma}   \\
0      &  -s_{23}e^{i\gamma} &   c_{23}
\end{array}  \right)
\left(
\begin{array}{ccc}
c_{13}             &   0    &   s_{13}e^{-i\beta}     \\
0                  &   1    &   0                     \\
-s_{13}e^{i\beta}  &   0    &   c_{13}
\end{array}  \right)
\left(
\begin{array}{ccc}
c_{12}                &  s_{12}e^{i\alpha}  & 0       \\
-s_{12}e^{-i\alpha}   &  c_{12}             & 0       \\
0                     &  0                  & 1  
\end{array}  \right)    ,
\end{equation} 
where $c_{ij}\equiv\cos\psi_{ij},s_{ij}\equiv\sin\psi_{ij}$
with $\alpha,\beta$ and $\gamma$ denoting the dynamical CP phases in matter.
From the off-diagonal elements of Eq.~(\ref{dyn}) we can obtain the following
differential equations for three mixing angles and three mixing phases,
\begin{equation}
\left(
\begin{array}{c}
\dot{\psi}_{12}  \\
\dot{\psi}_{13}  \\
\dot{\psi}_{23}  \\
\dot{\alpha}     \\
\dot{\beta}      \\
\dot{\gamma}
\end{array}     \right)
=\frac{1}{2E}{\cal B}^{-1}
\left(
\begin{array}{c}
\hbox{Re}(U^{\dagger}_{dyn}M^wU_{dyn})_{12}       \\
\hbox{Im}(U^{\dagger}_{dyn}M^wU_{dyn})_{12}       \\
\hbox{Re}(U^{\dagger}_{dyn}M^wU_{dyn})_{13}       \\
\hbox{Im}(U^{\dagger}_{dyn}M^wU_{dyn})_{13}       \\
\hbox{Re}(U^{\dagger}_{dyn}M^wU_{dyn})_{23}       \\
\hbox{Im}(U^{\dagger}_{dyn}M^wU_{dyn})_{23}       
\end{array} \right)          .
\label{eq1}
\end{equation}
Here the $6\times 6$ matrix ${\cal B}$ is given by
\begin{equation}
{\cal B}=\left(
\begin{array}{cccccc}
-s_{\alpha}  & 0  &  
    c^2_{12}s_{13}s_{\beta-\gamma}-s^2_{12}s_{13}s_{2\alpha+\beta-\gamma} &
   -s_{12}c_{12}c_{\alpha}     &    -s_{12}c_{12}s^2_{13}c_{\alpha}  &  b_{16}  \\
 c_{\alpha}  & 0  &  
    c^2_{12}s_{13}c_{\beta-\gamma}+s^2_{12}s_{13}c_{2\alpha+\beta-\gamma} &
   -s_{12}c_{12}s_{\alpha}     &    -s_{12}c_{12}s^2_{13}s_{\alpha}  &  b_{26}   \\
0  &  c_{12}s_{\beta}  &  s_{12}c_{13}s_{\alpha-\gamma}  & 
    0        &  c_{12}s_{13}c_{13}c_{\beta}                          &  b_{36}    \\
0  &  c_{12}c_{\beta}  &  -s_{12}c_{13}c_{\alpha-\gamma}  &
    0        &  -c_{12}s_{13}c_{13}s_{\beta}                          &  b_{46}    \\
c_{12}c_{13}s_{\gamma}  &  s_{12}s_{\alpha+\beta}  &  0     &
    0        &  s_{12}s_{13}c_{13}c_{\alpha+\beta}                    &  b_{56}    \\
c_{12}c_{13}c_{\gamma}  &  s_{12}c_{\alpha+\beta}  &  0     &
    0        &  -s_{12}s_{13}c_{13}s_{\alpha+\beta}                    &  b_{66}
\end{array}    \right)    ,
\end{equation}
where
\begin{eqnarray}
 b_{16} &=& s_{12}c_{12}(1+s^2_{13})s^2_{23}c_{\alpha} +
            s^2_{12}s_{13}s_{23}c_{23}c_{2\alpha+\beta-\gamma} -
            c^2_{12}s_{13}s_{23}c_{23}c_{\beta-\gamma}      ,    \nonumber   \\
 b_{26} &=& s_{12}c_{12}(1+s^2_{13})s^2_{23}s_{\alpha} + 
            s^2_{12}s_{13}s_{23}c_{23}s_{2\alpha+\beta-\gamma} +
            c^2_{12}s_{13}s_{23}c_{23}s_{\beta-\gamma}       ,   \nonumber   \\ 
 b_{36} &=& -s_{12}c_{13}s_{23}c_{23}c_{\alpha-\gamma} -
              c_{12}s_{13}c_{13}s^2_{23}c_{\beta}             ,  \nonumber  \\
 b_{46} &=& -s_{12}c_{13}s_{23}c_{23}s_{\alpha-\gamma} + 
              c_{12}s_{13}c_{13}s^2_{23}s_{\beta}              , \nonumber  \\ 
 b_{56} &=& -s_{12}s_{13}c_{13}s^2_{23}c_{\alpha+\beta} +
             c_{12}c_{13}s_{23}c_{23}c_{\gamma}               ,  \nonumber  \\
 b_{66} &=& s_{12}s_{13}c_{13}s^2_{23}s_{\alpha+\beta} -
             c_{12}c_{13}s_{23}c_{23}s_{\gamma}   .             \nonumber                   
\end{eqnarray}  
The energy eigenvalues are
\begin{eqnarray}
 {\cal E}_1(t) &=& (U^{\dagger}_{dyn}M^wU_{dyn})_{11}
    -2E\{\dot{\alpha}s^2_{12}-\dot{\beta}c^2_{12}s^2_{13}-              \\
  &  & \dot{\gamma}[(s^2_{12}-c^2_{12}s^2_{13})s^2_{23}-
       2s_{12}c_{12}s_{13}s_{23}c_{23}c_{\alpha+\beta-\gamma}]    
      -2\dot{\psi}_{23}s_{12}c_{12}s_{13}s_{\alpha+\beta-\gamma}\} , \nonumber  \\
 {\cal E}_2(t) &=& (U^{\dagger}_{dyn}M^wU_{dyn})_{22}            
    +2E\{\dot{\alpha}s^2_{12}+\dot{\beta}s^2_{12}s^2_{13}+          \nonumber  \\
  &  & \dot{\gamma}[(c^2_{12}-s^2_{12}s^2_{13})s^2_{23}-          
       s_{12}c_{12}s_{13}s_{23}c_{23}c_{\alpha+\beta-\gamma}]
     -\dot{\psi}_{23}s_{12}c_{12}s_{13}s_{\alpha+\beta-\gamma}\}   , \\
 {\cal E}_3(t) &=& (U^{\dagger}_{dyn}M^wU_{dyn})_{33}
    -2E\{\dot{\beta}s^2_{13}+\dot{\gamma}c^2_{13}s^2_{23}\} .
\label{eq2}
\end{eqnarray}
The averaged transition probability is of the following form:
\begin{equation}
<P(\nu_e\rightarrow\nu_e)>=\sum_{i=1}^{3}
               |(U_{dyn}^*)_{i1}(U_0)_{i1}|^2   .
\end{equation}
In the three generation case, the dynamical phases contribute
through $(U^*_{dyn})_{i1}$
to the transition probability in contrast to the two generation case.
The formulas in the three generation case are reduced 
to the corresponding expressions in the two generation case, 
when we take $\theta_{13}=\theta_{23}=0$
and $m_{3i}=m_{i3}=0 (i=1,2,3)$. 
In the case of three generation of neutrinos, however,
it is prohibitively difficult to solve the coupled
differential equation~(\ref{eq1})-~(\ref{eq2}) numerically
and such a study is beyond the scope of the present work.

\section{Conclusion}
In this paper 
we have introduced the effective mixing matrix that relates weak eigenstates
to mass eigenstates, and the dynamical mixing matrix that connects  
weak eigenstates and energy eigenstates.
Using these definitions, we have studied the effective mixing matrices 
for the two and three flavor mixing cases,
with emphasis on the nature and behavior of the CP phases in matter. 
It has been shown that in the case of three generation mixing, 
two additional induced phases appear when 
a vacuum CP phase is non-zero.
Analysis of the dynamical mixing matrices 
for the two and three flavor mixing cases has also been presented.
In contrast to the effective mixing matrix of the two flavor mixing, 
in the dynamical mixing matrix, it is necessary to introduce 
one additional phase in a medium in order to 
be self consistent.  
This does not depend on the neutrino type. 
We have used the solar neutrinos to demonstrate the 
differences involved in the behavior of the CP phases in matter,
and the differences between the effective and dynamical mixing matrices. 

\section*{Acknowlegments}
This work was supported in part by the 
Korea Science and Engineering Foundation (KOSEF)
and in part by the Center for Theoretical Physics, Seoul National University, Korea.
One of the authors (CWK) would like to thank Department of Physics, Korea
Advanced Institute of Science and Technology for the hospitality
extended to him while this work was in progress.

\newpage
\centerline{\bf Figure Captions}
\begin{description}
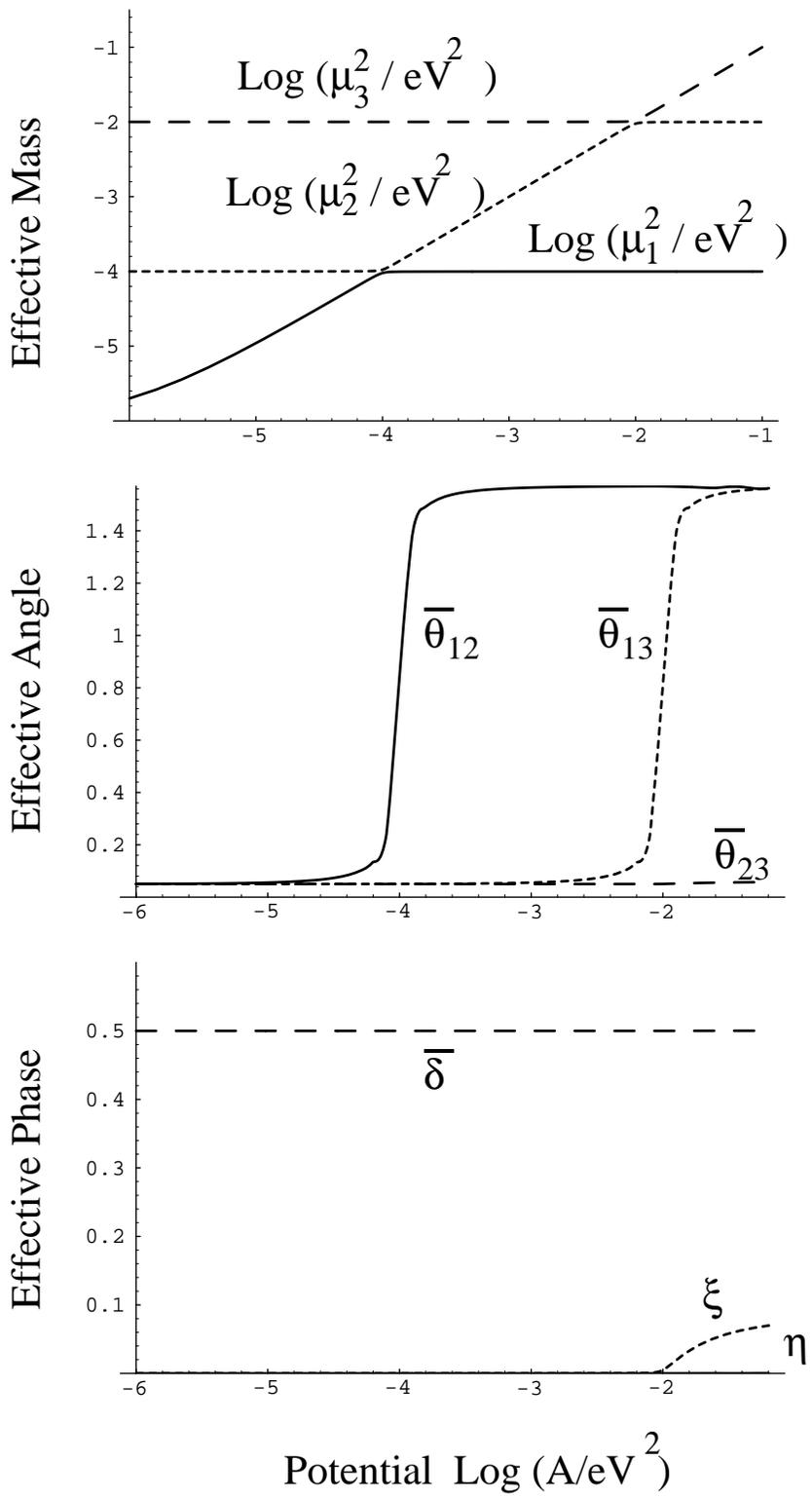
\item[Fig. 1]
\label{fig1}
The effective masses, mixing angles, and CP phases 
in matter as functions of $A$ with 
$m_1^2=10^{-6} \hbox{eV}^2,m_2^2=10^{-4} \hbox{eV}^2,
m_3^2=10^{-2} \hbox{eV}^2$, mixing angles
$\theta_{12}=\theta_{13}=\theta_{23}=0.05$ and 
vacuum CP phase $\delta=0.5$.
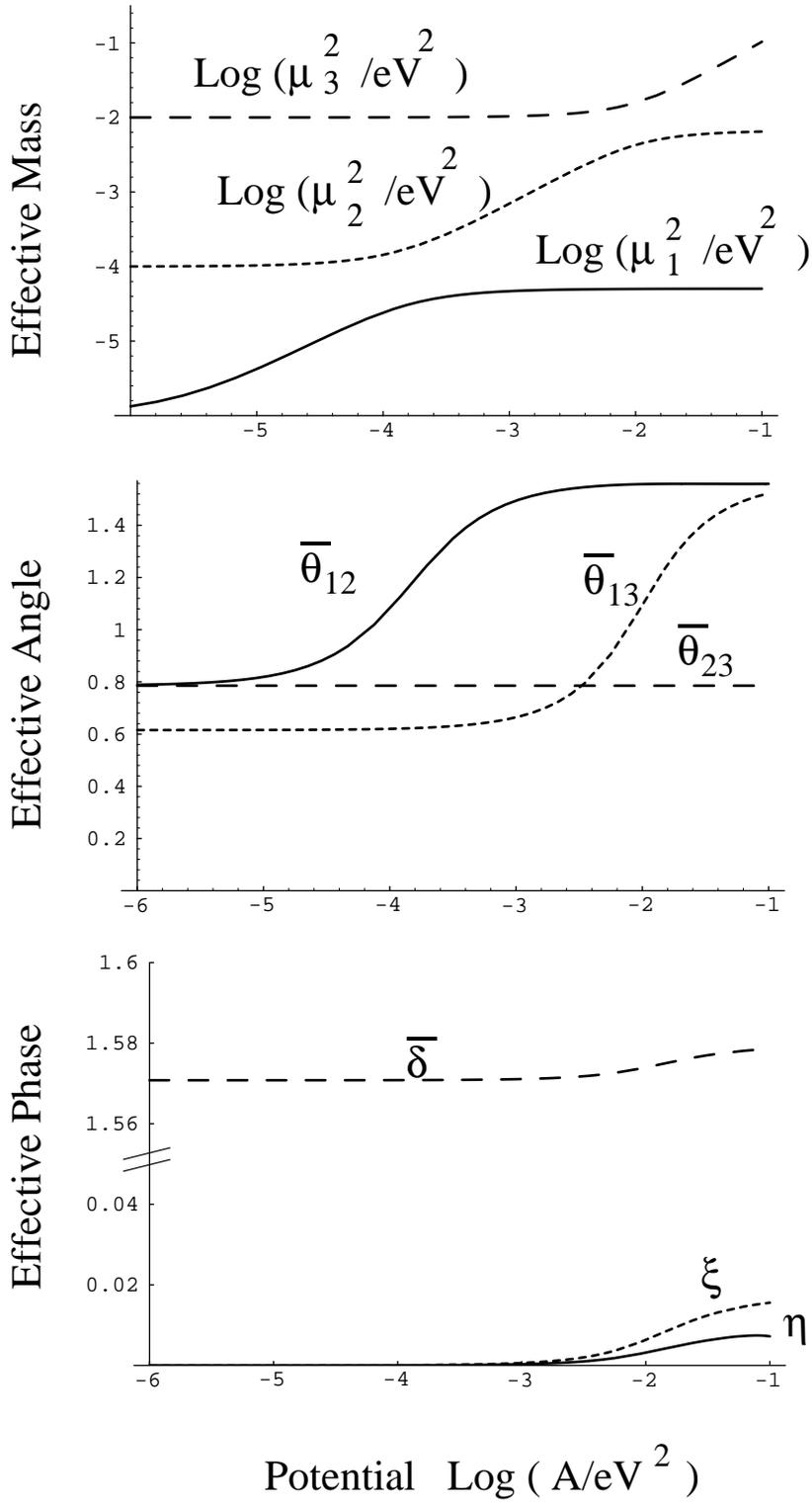
\item[Fig. 2]
The effective masses, mixing angles, and CP phases 
in matter as functions of $A$ for the maximal mixing case: 
$m_1^2=10^{-6} \hbox{eV}^2,m_2^2=10^{-4} \hbox{eV}^2,
m_3^2=10^{-2} \hbox{eV}^2$,vacuum mixing angle
$\theta_{12}=\pi/4,\theta_{13}=0.6155,\theta_{23}=\pi/4$,
and vacuum CP phase $\delta=\pi/2$. 
\label{fig2}
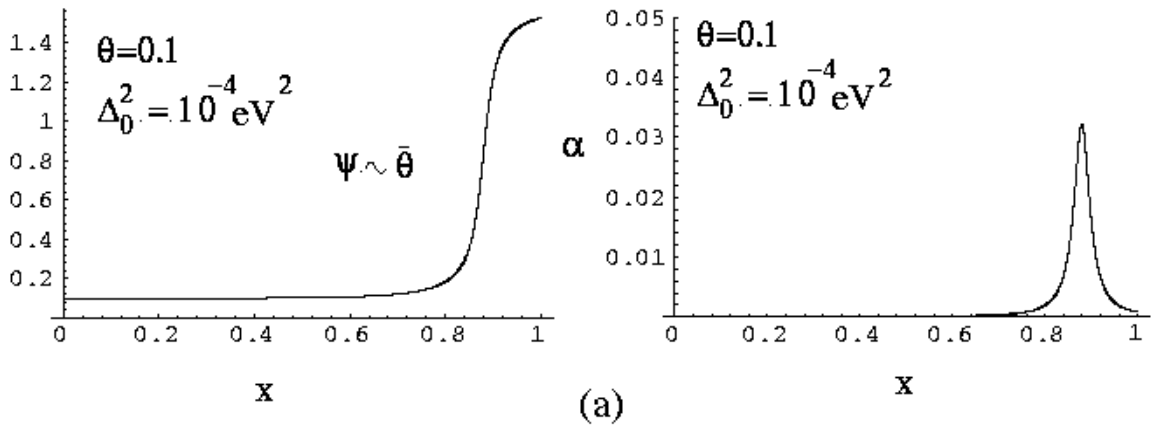
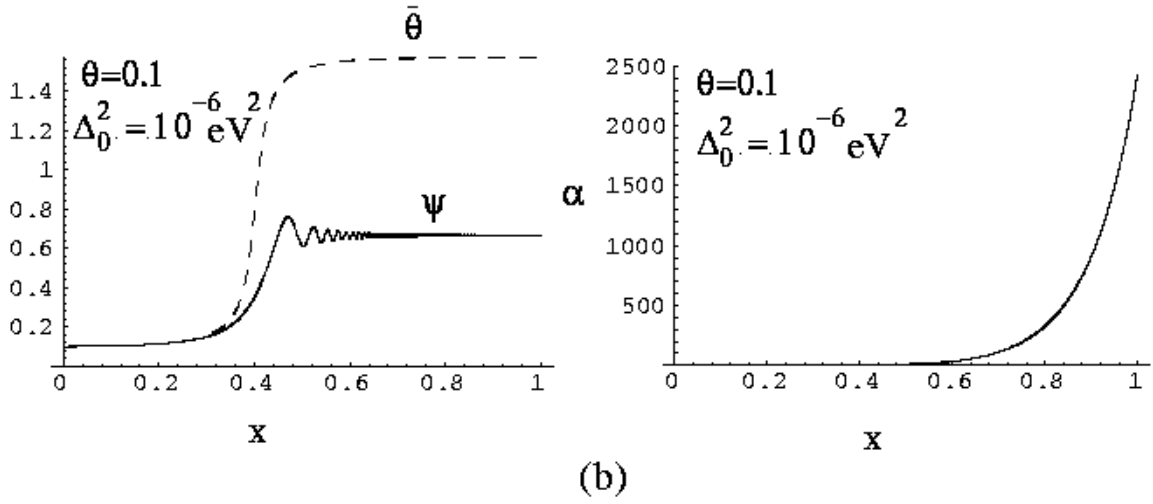
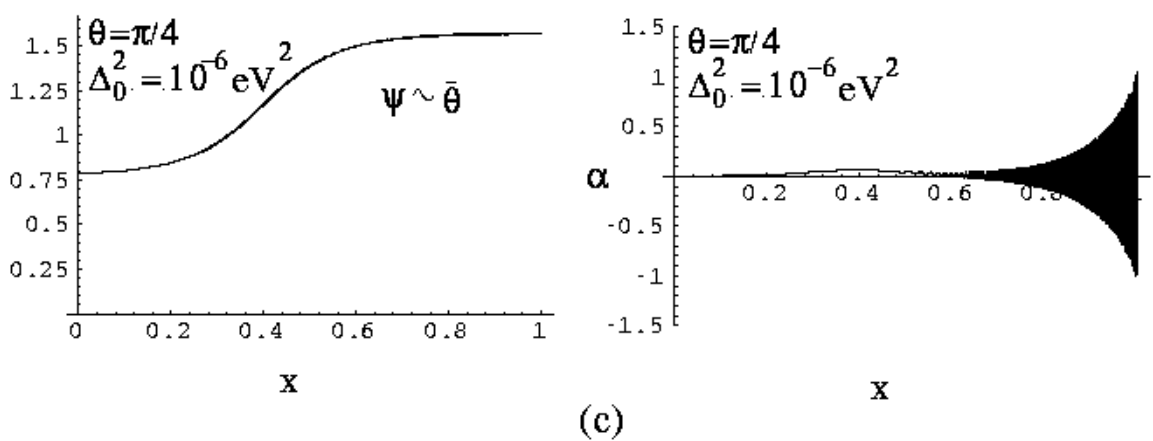
\item[Fig. 3]
The effective ($\bar\theta$) and dynamical ($\psi$) mixing angles and 
the dynamical mixing phase, $\alpha$ in the case of
two generations. The parameters used are
$E_{\nu}=10$ MeV with
(a) the vacuum mixing angle $\theta=0.1$,
$\Delta^2_0=10^{-4}\hbox{eV}^2$, 
(b) the vacuum mixing angle $\theta=0.1$,
$\Delta^2_0=10^{-6}\hbox{eV}^2$, and
(c) the maximal mixing with two generations $\theta=\pi/2$,
$\Delta^2_0=10^{-6}\hbox{eV}^2$.
The abscissa $x$ indicates the ratio of distance from the solar surface 
to the radius of the Sun. The value $x=0$ means the surface
of the Sun and $x=1$ the center of the Sun.
The sign of the phase $\alpha$ is reversed when the neutrinos
travel inwards.
\label{fig3}
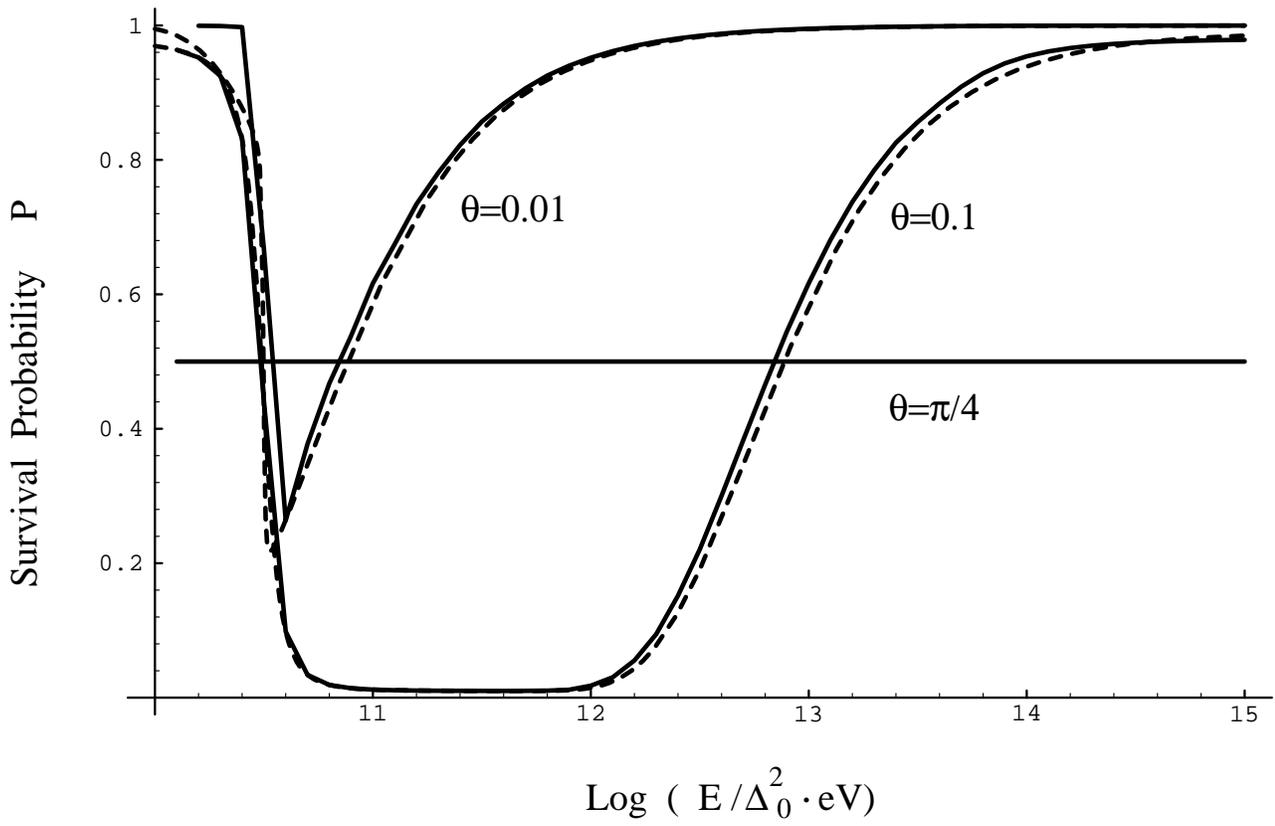
\item[Fig. 4] 
The $\nu_e$ survival probabilities in the two generation case as functions of
$E_{\nu}/\Delta^2_0$ for the vacuum mixing angle $\theta=0.01$, $\theta=0.1$,
and $\theta=\pi/4$ (maximal mixing).
The solid curves are the results  
with the use of the dynamical mixing angles whereas the dotted curves
($\theta=0.01$, $\theta=0.1$) are those with the use of the standard 
two generation formulas with the Landau-Zener transition probabilities.
The Landau-Zener formula cannot be applied to the 
case of large mixing, specifically to the maximal mixing mixing case.
\label{fig4}
\end{description}
\end{document}